\documentclass{PoS}
\usepackage{url}

\title{JLQCD IroIro++ lattice code on BG/Q}

\ShortTitle{JLQCD IroIro++ lattice code on BG/Q}

\author{\speaker{Guido Cossu}, Jun Noaki, Shoji Hashimoto, Takashi Kaneko\\
  KEK Theory Center, IPNS\\
  Tsukuba, Ibaraki 305-0810, Japan\\
  E-mail: \email{cossu@post.kek.jp, noaki@post.kek.jp, shoji.hashimoto@kek.jp, takashi.kaneko@kek.jp}}

\author{Hidenori Fukaya\\
  Department of Physics, Osaka University\\
  Toyonaka 560-0043, Japan\\
  E-mail: \email{hfukaya@het.phys.sci.osaka-u.ac.jp}}

\author{Peter A. Boyle\\
  University of Edinburgh\\
  Edinburgh, UK\\
  E-mail: \email{paboyle@ph.ed.ac.uk}}

\author{Jun Doi\\
  IBM Research, Tokyo, Japan\\
  E-mail: \email{doichan@jp.ibm.com}}

\abstract{We describe our experience on the multipurpose C++ code IroIro++ designed for JLQCD to run on the BG/Q installation at KEK. We discuss some details on the performance improvements specific for the IBM Blue Gene Q.}

\FullConference{31st International Symposium on Lattice Field Theory - LATTICE 2013\\
		July 29 - August 3, 2013\\
		Mainz, Germany}

\begin{document}

\section{Code structure overview}

Over the last few years the complexity of architectures and algorithms for lattice gauge theory simulations has got higher and higher at an unprecedented fast pace. The legacy FORTRAN code used at KEK since the acquisition of the new generation of machines (IBM Blue Gene/Q and Hitachi SR16K) could not deal anymore with such level of ramification. Maintainability and support for an heterogeneous set of platforms were hard to ensure. Before the arrival of the new generation of machines the JLQCD group decided to rewrite from scratch a code for lattice gauge theory simulations using the already available codes in the market as a source of inspiration for best practices. 

The essential design targets are:
\begin{itemize}
\item a complete abstraction of physical concepts from the actual implementation;
\item a user-level that does not need a single line of programming to change the code behavior;
\item maintainability of many code paths and sub-cases at the developer-level;
\item support for multiple platforms and optimization on some specific architectures.
\end{itemize}

The object orientation paradigm fits nicely such a complexity and, although creating initial performance issues, the C++ language was a natural choice for its flexibility. We could achieve the desired functionality using the design patterns programming paradigm \cite{DesignPatterns} influenced by the Chroma code-set \cite{Chroma}, but keeping code complexity at a manageable level.

The structure is layered. The communication level (could be single core, MPI, BGNET) is the lowest one. Several basic concepts are defined, like a Field (just a container) and Formats (that define the content of a Field), up to the definitions of Dirac operators, Actions, Solvers, Trajectories, and so on, as abstract concepts. Some libraries come with an architecture dependent optimized implementation, easily extendable to future designs.

The code is now mature, it is actually used intensively for configuration generation and analysis in our current $N_f=2+1$ Domain-Wall fermion runs, and it has been tested on various platforms including UNIX PC (single/multi-core), Mac, Blue Gene/Q, Hitachi SR16K. 

The performance optimization efforts were concentrated so far on the IBM Blue Gene/Q architecture, the biggest installation at KEK in terms of computational power. Special libraries were developed by IBM Japan to speed-up communications, the Dirac operator kernels and the linear algebra \cite{IBM}, see section \ref{sect:BGQ} for more details. On top of that we implemented a full support for the highly optimized kernels of BFM \cite{Boyle:2009vp} to further increase performance, see section~\ref{sect:BFM}. An improvement on the performances on the Hitachi SR16K (IBM Power7 architecture) is currently in progress.

The list of features of the current (October 2013) implementation includes:
\begin{itemize}
\item  Actions (Gauge: Wilson, Rectangle, Fermion: 2 flavors, 2 flavors ratio of determinants (Hasenbusch preconditioning~\cite{Hasenbusch:2001ne}), RHMC $N_f$ flavors, RHMC $N_f$ flavors ratio of determinants, Overlap)
\item  Dirac operators (Wilson, Clover, Staggered, Adjoint Staggered, Overlap, Even-odd preconditioned Wilson, Generalized Domain Wall (4d-5d), Wilson Brillouin, Hybrid DomainWall-Overlap, M\"obius Kernel)
\item  Linear Solvers (Conjugate Gradient, BiCG Stabilized, Rational-Multishift)
\item  Measurements 
  \begin{enumerate}
  \item Quark propagators [Wilson, Domain Wall], momentum space propagators, Meson and Baryon correlators, Eigenmodes, Low mode preconditioning, Gauge fixing - Coulomb \& Landau,
  \item Gauge quantities [Plaquette, Polyakov Loop, Wilson Loop], Topological charge, Wilson Flow \cite{Luscher:2010iy}
  \end{enumerate}
\item  Smearing (APE, Stout analytic \cite{Morningstar:2003gk}), HMC runs with smeared fermionic actions
\item  Random Number Generators (Mersenne Twister \cite{MT}, Dynamic Creation Mersenne Twister \cite{MT})
\item  Full I/O support for plain ASCII and plain binary, just reading support for ILDG, NERSC, MILC, JLQCD-legacy type of configurations
\item  BAGEL/BFM integration \cite{Boyle:2009vp}
\item  User-level XML control of program behavior 
\end{itemize}

IroIro++ is currently under active development and the list of supported features is changing month by month. Several things are still to be completed before a full public release. The code still lacks an organized and robust test suite that is necessary to keep track efficiently of regression bugs. This is a point to be addressed in the near future. Unfortunately documentation is still limited to the Doxygen files and some basic information on compilation and running simple jobs\footnote{See this documentation at the page: \url{http://suchix.kek.jp/guido_cossu/documents/DoxyGen/html/index.html}}.

In the following sections of this report we discuss the performance optimization related to the BG/Q architecture, first reviewing the libraries by IBM Japan and then the BFM integration. In the final section we present some performance figures comparing several setups.

\section{Blue Gene/Q and IBM libraries\label{sect:BGQ}}

The Blue Gene/Q is the third in line of the IBM Blue Gene series. It was designed for high scalability and excellent performance per watt which makes it currently one of the best choices for HPC.
Every board contains 32 chips running at 1.6 GHz and communicating through a metal plane. The nodes communicate through each other through a 5d network of optical cables. Every chip has 16+1+1 PowerPC A2 processor cores and applications can use 16 of them. Each processor supports 4-way Simultaneous Multi-Threading (SMT) for up to 64 concurring threads, and has also a 4-way SIMD FPU that implements fused multiply-add operations (FMA), thus giving a theoretical peak performance of 204.8 GFlops. Table~\ref{table:BGQ}  summarizes the technical details and also the hierarchical memory structure, emphasizing that a network bandwidth close to the DDR memory bandwidth guarantees the scalability up to sizes of 120 racks (more than 2 million cores, at LLNL).

\begin{table}[h]
\small
\centering
\begin{tabular}{ l  l }
\hline
Processor   &    IBM PowerPC A2 1.6 GHz, 16 cores per node \\
Main Memory &    16 GB SDRAM-DDR3 per node, 42.6 GB/s \\
L1 Cache    &    16 KB data, 820 GB/s \\
L2 Cache    &    32 MB per node, 563 GB/s\\
Network     &    5D Torus - 10 bi-directional links = 40 GBps; 2.5 $\mu$sec latency\\
Power       &    Typical 80 kW per rack (estimated)\\
\hline
\hline
\end{tabular}
\caption{Some of the specifications of the IBM Blue Gene/Q series. The importance of a code that uses efficiently the memory hierarchy is evident from the relative bandwidths.}
\label{table:BGQ}
\end{table}

\subsection{Communication and threading: the BGNET and the BGQThreads libraries}

IBM Japan developed a series of libraries (BGNET, BGWilson, BGQThreads) to exploit the computational capabilities of BG/Q. The BGNET library is a low latency, light-weight API with smaller overhead for small data packets than MPI. It is programmed using the System Programming Interface (SPI) APIs by IBM and it uses the remote DMA (rDMA) capabilities of communication between compute nodes on BG/Q. It supports asynchronous send and receive, small overhead barriers, global sum and reduce, broadcasts and more. IroIro++ fully supports this communication layer, as well as plain MPI, which is slower. The BGQThreads library is an SMP threading library that has lower latencies for barriers than the standard openMP functions. 

\subsection{Kernels}
\begin{figure}[t]
\centering
\includegraphics[clip=true,width=0.4\textwidth]{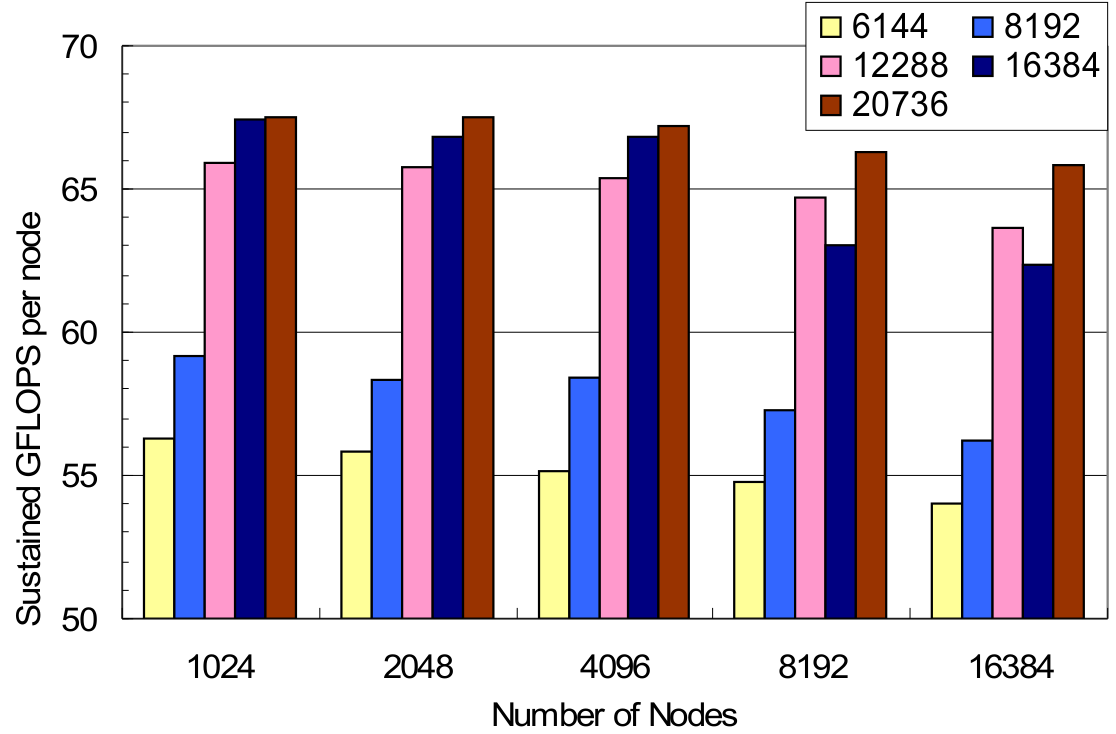}
\caption{(BGWilson BiCGStab solver) Comparison of the sustained GFlops/node using different number of sites for the local lattice (color index) and varying the total number of nodes. This shows excellent weak scaling of the BGWilson library solver. From~\cite{IBM}.}
\label{fig:weakscaling}
\end{figure}

The kernels written by J. Doi \cite{IBM} use an array of structures memory pattern for the field storage which has several advantages for cache memory access although it is not portable. 
BGWilson is a library for the Wilson Dirac matrix-vector multiplication (in chiral and Dirac representations) that is written in C using intrinsics for SIMD data loading and processing, optimization of the pipeline by assembly code, threading to hide latencies, concurrent data transfer and calculations and even-odd preconditioning. The library is extended to include full CG solvers for the Dirac operator. All the 5D Domain Wall kernels and solvers are optimized using functions from this library, minimizing the creations of temporary objects that typically affects performance on C++ codes. The measured performances for this library are about 65 GFlops per node for BiCG Wilson solver (see figure~\ref{fig:weakscaling} for a weak scaling test) and 25 GFlops per node for 5D Domain Wall fermion solver (in double precision). Since using a specialized solver breaks the rule of having distinct objects for different concepts (the Dirac operator and the solvers) we had to create special solvers classes that are used when the improved libraries are called. This translates at user-level in a simple change in the name of the fermionic action.

Besides the optimization of the Dirac kernels all the linear algebra for matrix-matrix, matrix-vector, vector-vector operations are threaded and use SIMD functionalities. They can be called from within an openMP parallel section dividing the workload in chunks to be processed by each single thread. In the current version of the code all the computational steps of the (R)HMC code use the hybrid MPI-openMP threading model, using all the level of parallelism that BG/Q provides (SIMD, in core-threading, inter-core communications).  In relation to performance we examine the balance between the number of MPI processes and the openMP threads in section~\ref{sect:Perf}.

\section{BFM: Bagel Fermion Matrix library integration\label{sect:BFM}}

\begin{figure}[b]
\centering
\includegraphics[clip=true,width=0.45\textwidth]{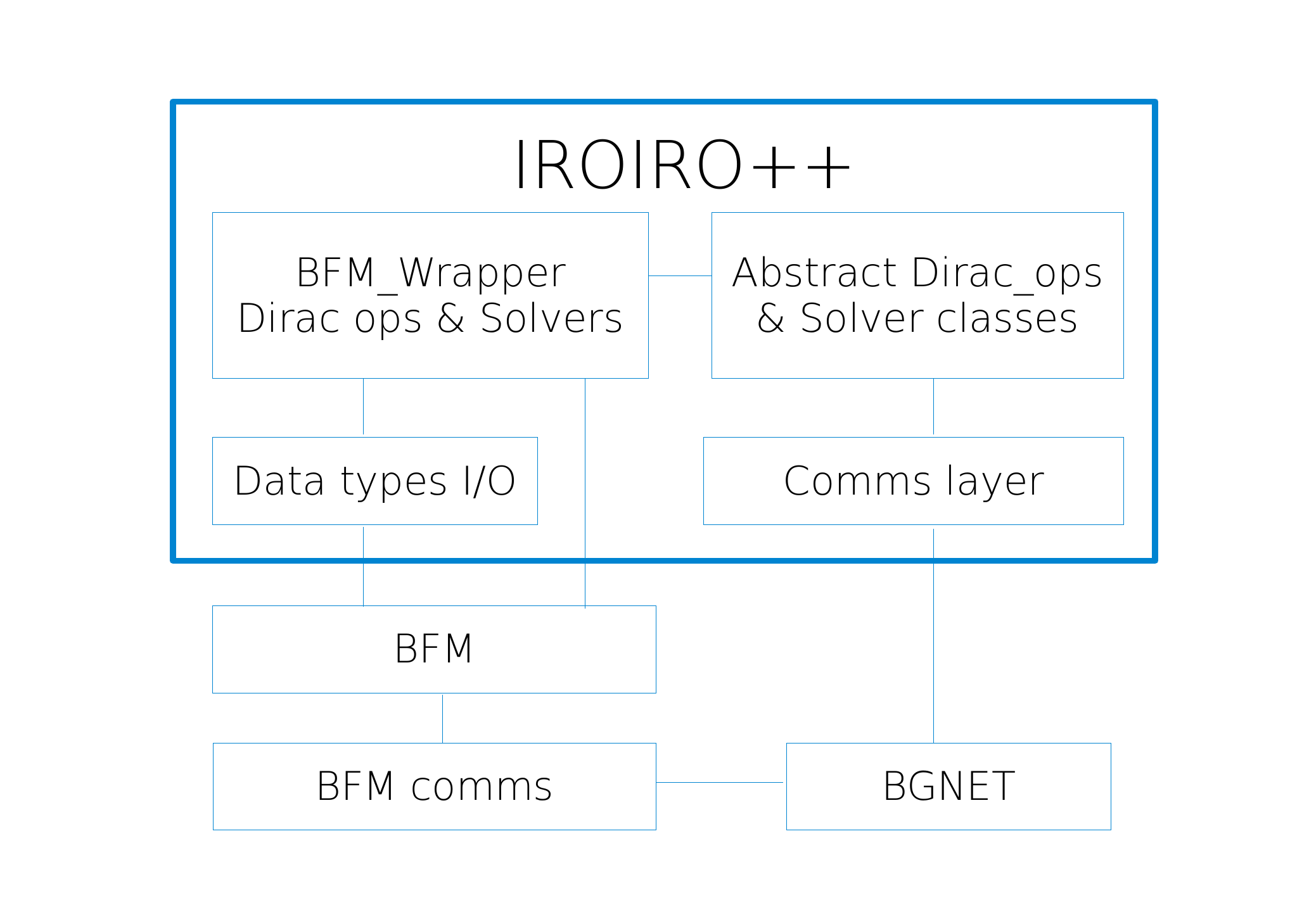}
\caption{Scheme of the dependencies for IroIro++ on the BFM libraries.}
\label{fig:IroIroBFM}
\end{figure}

Integrating the Bagel Fermion sparse-Matrix (BFM) library by P. Boyle \cite{Boyle:2009vp} led to another boost in the code efficiency. The Bagel package is a domain specific compiler targeted to several architectures that have been optimized for all the BlueGene series microprocessors (L, P, Q). It generates assembler code for the core routines of a QCD calculation starting from a table of mappings from abstract primitive operations to the actual architecture dependent instructions. Then the code performs efficient prefetching of data streams and rescheduling of the issued instructions. The low level assembly code is then used by the BFM package that provides a useful interface for complex code-sets like IroIro++ or Chroma. BFM contains utilities for memory management to allocate, import/export into the BFM memory storage pattern the fermion and gauge fields. Thus before using any function in BFM we wrote an interface layer that translates the conventions in the memory storage and fermions representation between IroIro++ and BFM (the current overhead for this step is about 1\%). As the BGWilson library, BFM contains already optimized CG solvers for Wilson and 5D Domain-Wall fermions. For communications, it relies on external libraries, in this case the BGNET. The scheme of dependencies of the IroIro++ code on the BFM library is depicted in figure~\ref{fig:IroIroBFM}.

\section{Performances and conclusions\label{sect:Perf}}

\begin{figure}[t]
\centering
\includegraphics[clip=true,width=0.55\textwidth]{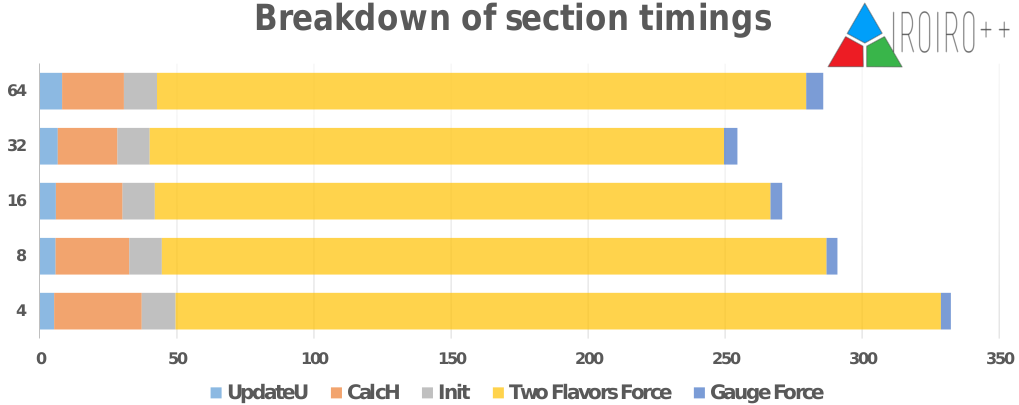}
\caption{Running times of two trajectories of $N_f=2$ Domain-Wall fermions, against the number of threads, see table~\protect\ref{table:perf1} for details. The fermion force calculation is not completely threaded yet.}
\label{fig:IroIroPerformance}
\end{figure}

In this section we report the current status regarding performances of IroIro++ in generating configurations using the (R)HMC algorithm by changing the balance between the number of threads and the number of MPI processes on BG/Q. We show in table~\ref{table:perf1} and figure~\ref{fig:IroIroPerformance} the results of a set of tests: two flavors of Domain Wall fermions (Scaled-Shamir kernel) keeping the global lattice size constant to $16^3\times 32$ and running on the smaller partition, 32 nodes. The product of MPI processes times the number of threads is kept constant to 64. In the last five columns we report the breakdown of the timings relative to the single sections of the HMC algorithm. The Gflops/node figure includes everything, code initialization included, so it is underestimated for such small number of trajectories. The parameters have not been chosen to get the maximum sustained performance that is around 30 GFlops/s per node. 

\begin{table}[h]
\small
\centering
\begin{tabular}{c|c|c|c|c||c|c|c|c|c}
\hline
N. Threads & Total & CG & \% in CG & GFlops/s & UpdateU & CalcH & Init & P-0 & P-1 \\
\hline
4 &       333 & 303 & 92.71  &  20.88 & 5.19   & 31.92  & 12.36 & 279.29 & 3.60 \\
8 &       292 & 258 & 90.10  &  24.28 & 5.70   & 26.90  & 11.88 & 242.52 & 4.00 \\
16 &       271 & 231 & 87.00 &  27.56 & 5.83   & 24.28  & 11.81 & 224.65 & 4.30 \\
32 &       255 & 202 & 81.09 &  27.35 & 6.49   & 21.75  & 11.81 & 209.56 & 4.86 \\
64 &       287 & 202 & 72.12 &  25.00 & 8.10   & 22.56  & 12.04 & 236.90 & 6.22 \\
\hline
\end{tabular}
\caption{Current running time (sec) and performance (GFlops per node) in a run using BFM on a 32 node partition, no fiber optics communications, using node maps. Global lattice size $16^3\times 32$, the mass $am=0.009$, the fifth dimension $L_s=12$. The 4th column show the time spent in CG relative to the whole HMC trajectory. UpdateU stands for the exponentiation part timing in seconds, CalcH for the calculation of the action at the beginning and at the end of the trajectory, Init for the random numbers and gauge initialization and P-\# are the force calculation timings for the fermionic and the gauge part respectively. Summing-up all these contribution gives the total running time, that is for two trajectories with $\tau=1$ and 8 steps of molecular dynamics. Gauge is integrated in a 4 times finer interval.}
\label{table:perf1}
\end{table}

In table~\ref{table:perf2} a more realistic RHMC run on a 2048 nodes partitions is presented. The lattice is now $32^3\times 64$, simulating 2+1 flavors ($m_{u.d}= 0.007$, $m_s = 0.04$) with a Hasenbusch preconditioning step ($am=0.04$). The force terms are now labeling the preconditioner, the strange quark, the light quarks, and the gauge part respectively. The performance here suffer from the communications and synchronization among many cores and the incomplete threading of some parts of the code: the best overall is $\sim$15 GFlops/node (that drops to 13 GFlops/node if I/O is taken into account), while the CG alone is over 20 GFlops/node. 

\begin{table}[h]
\small
\centering
\begin{tabular}{c|c|c|c||c|c|c|c|c|c|c}
\hline
N. Threads & Total & CG & \% in CG & UpdU & CalcH & Init & P-0 & P-1 & P-2 & P-3\\
\hline
16 &       109 & 77 & 70.31   & 0.64   & 15.37  & 2.71 & 25.50 & 43.54 & 21.28& 0.51 \\
32 &       123 & 69 & 56.64   & 0.73   & 14.15  & 2.67 & 24.64 & 58.52 & 21.94 & 0.59 \\
64 &       185 & 70 & 38.16   & 0.93   & 14.77  & 3.07 & 26.97 & 108.67 & 29.69 & 0.70 \\
\hline
\end{tabular}
\caption{Timings for one trajectory on a 2048 nodes partition, as explained in the text. Labels are as in table~\protect\ref{table:perf1}, P-\# are respectively the force terms for: Hasenbusch term, Strange quark (rational approximation), Two flavors, Gauge.}
\label{table:perf2}
\end{table}

The sweetest point for CG is always when the number of threads is maximum, so we can avoid some of the expensive MPI communications and barriers. Most of the code is correctly threaded, showing almost no dependence, or a mild one, on the number of threads. The missing part is the force calculation that has a severe impact on the final performance, especially when the rational approximation of the force terms is used for the strange quark (P-1 in table~\ref{table:perf2}). This improvement is currently under development at the time of writing. 

In conclusion, we presented our efforts in writing an efficient but easily extendable code for lattice simulations on big scale supercomputers. The code needs few more optimization to reach the optimal performance for every case (the target being the BFM CG efficiency).
The IroIro++ code is now almost ready for being published online, as soon as sufficient documentation is written to support new users.  

We acknowledge the people from the Bridge++ \cite{Bridge} group for their collaboration in earlier stages of this project.
Numerical simulations are performed on the IBM System Blue Gene Solution at High Energy Accelerator Research Organization (KEK)
under a support of its Large Scale Simulation Program (No. 12/13-04). This work is supported in part by the SPIRE (Strategic Program for Innovative REsearch) Field5 project, and H.F. is supported by the Grant-in-Aid of the Japanese Ministry of Education (No. 25800147).

\end{document}